 \documentclass[]{article}

\usepackage{arxiv}
\usepackage{setspace}
\usepackage{natbib}
\usepackage{moreverb,url}

\usepackage[colorlinks,bookmarksopen,bookmarksnumbered,citecolor=red,urlcolor=red]{hyperref}
\usepackage{soul}
\newcommand{\btheta}{{\pmb \theta}}

\newcommand\BibTeX{{\rmfamily B\kern-.05em \textsc{i\kern-.025em b}\kern-.08em
		T\kern-.1667em\lower.7ex\hbox{E}\kern-.125emX}}

\newtheorem{proposition}{Proposition}[section]

\newtheorem{remark}{Remark}[section]

\usepackage[T1]{fontenc}
\usepackage{hyperref}
\usepackage[latin9]{inputenc}
\usepackage{geometry}
\usepackage{array}
\usepackage{units}
\usepackage{pdfpages}
\usepackage{multirow}
\usepackage{amsmath}
\usepackage{amssymb}
\usepackage{cancel}
\usepackage{graphicx}
\PassOptionsToPackage{normalem}{ulem}
\usepackage{ulem}

\usepackage{amsfonts}       % blackboard math symbols
\usepackage{nicefrac}       % compact symbols for 1/2, etc.
\usepackage{microtype}      % microtypography
\usepackage{lipsum}

\usepackage{xcolor}

\begin{document}

\title{Quantifying replicability and consistency in systematic reviews}

	\author{
		Iman Jaljuli \\%\thanks{....} \\
		Department of Statistics and Operations Research, \\
		Tel-Aviv University, \\
		Tel-Aviv, Israel.\\
		\texttt{jaljuli.iman@gmail.com} \\
		\And
		Yoav Benjamini \\% \thanks{...} \\
		Department of Statistics and Operations Research, \\
		Tel-Aviv University, \\
		Tel-Aviv, Israel.\\
		\texttt{ybenja@gmail.com} \\
		\And
		Liat Shenhav \\% \thanks{...} \\
		Department of Computer Science, \\
		University of California Los Angeles, \\
		Los Angeles, CA, USA. \\
		\texttt{liashenhav@gmail.com} \\	
		\And
		Orestis Panagiotou \\ %\thanks{...} \\
		Department of Health Services, Policy \& Practice, \\
		Brown University, \\
		USA.\\
		\texttt{orestis\_panagiotou@brown.edu} \\	
		\And
		Ruth Heller \\ %
		Department of Statistics and Operations Research, \\
		Tel-Aviv University, \\
		Tel-Aviv, Israel.\\
		\texttt{ruheller@gmail.com} \\
	}

\maketitle

\begin{abstract}
{Systematic  reviews are important tools for synthesizing evidence from multiple studies. They serve to increase power and improve precision, in the same way that larger studies can do, but also to establish the consistency of effects and replicability of results across studies which are not identical. In this work we propose to incorporate tools to quantify \emph{replicability} of  effects signs (or directions).
 We suggest that these tools accompany  the fixed-effect or random-effects meta-analysis,
 and we show that they convey important additional information for the assessment of the intervention under investigation. 
 We motivate and demonstrate our approach and its implications by examples from systematic reviews from the Cochrane  library, and offer a way to incorporate our suggestions in their standard reporting system. Our tools make no assumptions on the distribution of the true effect sizes, so their inferential guarantees continue to hold even if the fixed-effect or random-effects model is false.}
{Cochrane Collaboration, Heterogeneity, Partial conjunction analysis, meta-analysis, $r$-value.}
\end{abstract}

\section{Introduction} \label{intro}

	In systematic reviews, several studies that examine the same question are analyzed together. Viewing all the available information is extremely valuable for practitioners in the health sciences. A notable example is the Cochrane systematic reviews on the effects of healthcare interventions \citep{Higgins11}. 	
Deriving conclusions about the overall health benefits or harms from an ensemble of studies can be difficult, since the studies are never exactly the same and there is danger that the differences between studies affect the inference.  

There are many reasons to perform a meta-analysis, according to the Cochrane Handbook for Systematic Reviews of Interventions (\S~9.1.3,
 \cite{Deeks19}). The first two reasons are the obvious ones: to increase power and improve precision.
The next two reasons
come to answer questions that cannot be addressed by  individual studies. We quote: (3) ''Primary studies often involve a specific type of patient and explicitly defined interventions. A selection of studies in which these characteristics differ can allow investigation of the consistency of effect and, if relevant, allow reasons for differences in effect estimates to be investigated."; (4) ''To settle controversies arising from apparently conflicting studies or to generate new hypotheses. Statistical analysis of findings allows the degree of conflict to be assessed formally, and reasons for different results to be explored and quantified."

Goals (3) and (4) are directly related to the growing concerns in recent years about lack of replicability of results in medical research and in science at large \citep{nuzzo2014scientific,mcnutt2014journals,collins2014policy}. The discussions are about the issues that may hurt the published research of a study, e.g.,  publication bias and unreported exploratory steps.
Most of the contemplated solutions involve the single stand-alone study: its design, pre registration, conduct, analysis and report, all in a reproducible way. 

What it means to have a result replicated  is not well established. Relying on \cite{fisher1935design}, a result has been replicated if the $p$-value is lower than some small threshold in both the original and the replicating studies (this was  Fisher's motivation for introducing a threshold for the $p$-value). This definition was used by "The Psychology Reproducibility Project" \citep{open2015estimating} and is still the most acceptable criterion to date.
One negative aspect of this approach is the emergence of the replicability effort as a single one-shot effort, ending with a clear conclusion of 'replicate' versus 'not-replicated'. For this reason, a major concern in the design of replication studies is to guarantee a large enough sample size, in order to assure sufficient power for making such a conclusion, see for example  report by the \cite{national2019reproducibility}. 
 
Systematic reviews offer a natural approach to assessing replicability, as reflected in Goal (3) where replication efforts are conducted only if the original result was of interest to other researchers. The studies are conducted by independent groups of researchers that try to follow similar protocols, but local deviations are unavoidable. Obviously, the different studies enlist different subjects, but furthermore they are often conducted on different populations from which subjects are drawn, sometimes considerably so. The individual studies are  also not necessarily of sufficient power, so that if they do not get a statistically significant result it cannot be concluded that there is a replicability problem with the original result. Nevertheless, the evidence from a number of small studies can be combined to assess whether the intervention effect has been replicated. 
This approach has been offered and replicability analysis tools were developed in a series of works by \cite{conj,Benjamini09,Heller10,wang2019admissibility}. The fourth goal goes beyond assessing replicability, in trying to identify not only lack of replicability but inconsistency of results, and then possibly explain their sources.

Traditionally in meta-analysis, such questions are addressed by the many methods to assess effect heterogeneity
\citep{higgins2002quantifying,Deeks19, panagiotou2015effect,borenstein2017basics,riley2011interpretation}. A widely used measure is $I^2 = \max\{ 0 , (Q-(n-1))/Q\}$, where $n$ is the number of studies, and $Q$ is the weighted sum of squared deviations from weighted mean across studies, with weights inversely proportional to the estimated study variances \citep{higgins2002quantifying,higgins2009re}. However,  the presence of effect heterogeneity  tells us nothing about whether the evidence is consistently in favour or against the intervention. 
Heterogeneity may be accompanied with high replicability, where all study effects may be positive (or all negative) but the confidence intervals ($CI$s) hardly overlap, such that the random effects (RE) model has low power and the model $p-$value is likely to be $>5\%$.  
It may also be accompanied with inconsistency, where the intervention is beneficial in some studies and harmful in others.

Another tool to monitor the consistency of an effect is sensitivity analysis, in which the meta-analysis is repeated, each time omitting one of the studies \citep{AnzuresCarbera10}.
The output is a plot of the results of these meta-analyses, called an "exclusion sensitivity plot" in \cite{Bax06},
 or \textit{leave-one-out analysis}  in \cite{JAMA_ioi200073}.

\begin{figure}[ht]
	\centering
	\includegraphics[height=7.5cm]{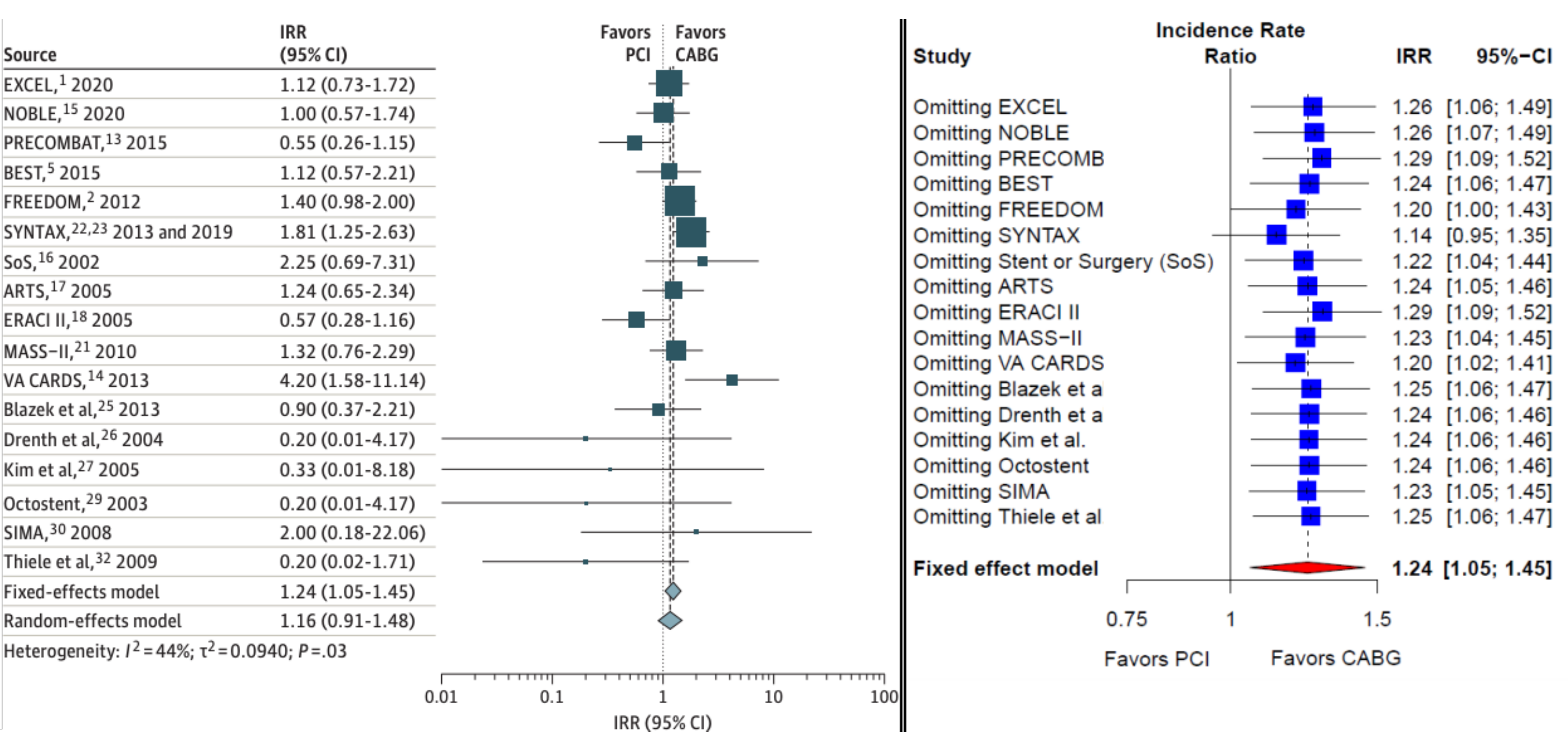}
	\caption{On the left-hand panel - Figure 2 in \citep{JAMA_ioi200073}.  On the right-hand panel - eFigure 5(A.) in their supplementary material.  }\label{fig-JAMA}
\end{figure}

\cite{JAMA_ioi200073}'s second figure is a forest plot
for the analysis of 17 studies (fig. \ref{fig-JAMA}, left panel).
The inference was based on the fixed-effects model due to the \textit{low heterogeneity}. The choice was further justified by an exclusion sensitivity plot (fig.~\ref{fig-JAMA}, right panel)  where they report that ''\textit{sensitivity analyses confirmed the solidity of the primary analysis}". However, this plot shows that the significance of the pooled estimated effect hinges solely on the results of SYNTAX: the single analysis omitting it has a corresponding overall estimate that is statistically insignificant, whereas each of the remaining 16 analyses (that include the study SYNTAX) unanimously declare significant findings. This clearly proves that  significance of the meta-analysis estimate hinges on the SYNTAX study  that steers the final result.

In this work we argue that in addition to reporting the estimated effect heterogeneity of the meta-analysis, it is important to quantify the consistency of evidence in favour or against the intervention, towards answering the goals in (3) and (4). 
We propose an inference method that 
quantifies effects replicability and  consistency. It allows us to make statements such as 
''\textit{with 95\% confidence, out of the five studies at least two studies have a positive effect}"
or \textit{''at least one study  has a negative effect and at least one study has an positive effect"}, see \S~\ref{sec-examples}.
Regarding the analysis by \cite{JAMA_ioi200073} displayed in figure \ref{fig-JAMA}, for example leads to "\textit{with 95\% confidence, out of the 17 studies at least one study has a positive effect}", but not the strong statement that there are at least two studies with positive effect. This statistical analysis indicates that the significance of the overall point estimate relies on a single study with no proven replicability.

Meta-analyses are mainly performed to estimate an overall treatment effect from the group of relevant studies \citep{Borenstein09} using the fixed effects (FE) or the random effects (RE) models, and the approach we offer is adopted to both. 
The models differ in their assumption about effects heterogeneity, where
the FE model assumes there is a common effect across studies. 
If this assumption is true, the test of the null hypothesis of no treatment effect  may be far better powered than the test in each of the individual studies.

However, if in fact there was heterogeneity in effect direction, then the RE model is recommended since it assumes the treatment effects are different, though they are also assumed to be independent identically distributed sample from a distribution (usually Gaussian). 
Some researchers argue that the fixed effect assumptions are implausible most of the time, and thus suggest to always use the RE model \citep{Higgins11,panagiotou2015effect},
where others make a choice based on clinical knowledge or based on a heterogeneity summary statistic. The Cochrane Handbook for Systematic Reviews \& Interventions \S~9.5.4 \citep{Deeks19} 
cautions against choosing a RE over FE meta-analysis based on a statistical test for heterogeneity. It is relevant to note that the RE model was also recommended by  \cite{Kafkafi05}   as the tool for assessing replicability across laboratories in animal phenotyping experiments.
A major criticism of the RE model is that validating the  distributional assumptions on the treatment effects is difficult \citep{Deeks19} and
it is not generally possible to distinguish whether heterogeneity results from clinical or methodological variability.

In \S~\ref{sec-rep}, we explain our methodology for quantifying the evidence towards consistency in effect direction that
tailors the existing general replicability analysis tools 
to meta-analyses in systematic reviews. 
In \S~\ref{sec-simulation} we demonstrate via simulations that the available meta-analysis tools do not provide the understanding our replicability analysis does. 
In \S~\ref{sec-examples} we demonstrate how such an evaluation contributes to the assessment of the intervention effects in case studies from the Cochrane library. 
In \S~\ref{sec-extentOfproblem} we summarize our evaluation of the extent of  consistency (and inconsistency) in effect direction  in the entire breast cancer domain of the Cochrane library. The special case of  the common-effect assumption is discussed  \S~\ref{sec-rvalueFE}; 
\S~\ref{sec-discuss} we discuss the interpretations of high heterogeneity and  conclude with some final remarks. 
We end this section with a brief review of the typical methods for meta-analysis carried out in Cochrane systematic reviews.

\section{ Replicability Analysis }\label{sec-rep}

Replicability of an effect size is usually treated as a virtue of the estimation process, however, it is more often than not a matter of the true effects estimated in the different studies. Hinging on this perception, we define replicability in terms of the true yet unknown parameters and build on these definitions to suggest a method for establishing replicability based on summery statistics.

\subsection{Replicability Framework}\label{subsec-frame}

Let $n$ be the number of studies available for meta-analysis,
$\btheta = (\theta_1,\ldots,\theta_n)$ the unknown treatment effects vector. 
For study $i\in \{1,\ldots,n\}$ 
the  null hypothesis of no treatment effect is, without loss of generality,  $H_i: \theta_i=0$. 
For convenience and practicality, assume no effect is zero effect. The effect is negative if $\theta_i<0$, the effect is positive if $\theta_i>0$.

For a group of $n$ studies and $u\in \{ 2,\ldots,n\}$, {\it $u/n$ replicability of a positive effect} is when
at least $u$ studies have a positive true effect, i.e., if the true effects vector $\btheta$ is in the set  
$$\mathcal A^{u/n}(R) = \left\lbrace\btheta: \sum_{i=1}^n I\left(\theta_i>0 \right)\geq u   \right\rbrace, $$ where $I(\cdot)$ is the indicator function. 
If $\btheta \in \mathcal A^{u/n}(R)$, at most $n-u$ studies have negative effects; otherwise, $n-u+1$ studies have negative effects. Similarly we define  {\it $u/n$ replicability of a negative effect} if $\btheta \in \mathcal A^{u/n}(L) = \left\lbrace\btheta: \sum_{i=1}^n I\left(\theta_i<0 \right)\geq u   \right\rbrace$.

Thus, we have {\it $u/n$ replicability} if at least $u$ studies have an effect in the same direction, $$\btheta\in \mathcal A^{u/n}(R)\cup \mathcal A^{u/n}(L). $$  
We define  {\it inconsistency} if 
  for $u\neq 0$, the {\it $u/n$ replicability} is met for both the {\it a positive } and  {\it a negative effect}. In particular,  {\it inconsistency}  is established if  {\it $1/n$ replicability}   is met for both a {\it positive} and a {\it negative effect}.

We have {\it consistency}  if at least two studies show an effect in the same direction, but no studies show an effect in the opposite direction, i.e., $\btheta$ is in the set  
$$ \left\lbrace \mathcal{A}^{2/n}(R)\backslash\mathcal{A}^{1/n}(L)\right\rbrace \cup\left\lbrace \mathcal{A}^{2/n}(L)\backslash\mathcal{A}^{1/n}(R)\right\rbrace $$

We aim to infer on a minimal number of parameters with true magnitude falling above and below 0. For establishing $u/n$ replicability, counting estimated parameters in each direction does not make valid inference.
In \S~\ref{subsec-NRhyp} and \S~\ref{subsecRV} we show how to establish $u/n$ replicability in each direction and overall, respectively.

\subsection{Establishing u/n replicability of effect direction}\label{subsec-NRhyp}
Let $\hat \theta_i,$ $\widehat{SE}_i$ be the estimated effect size and its standard error for study $i$.
For testing $H_i: \theta_i=0$, the test statistic  is $\hat \theta_i/ \widehat{SE}_i$ and the $p$-values for the left- and right-sided alternative hypotheses are $p_i^L, p_i^R$, respectively.

In order to establish $u/n$ replicability of a positive effects using the fore mentioned summary statistics, we test the composite null hypothesis that at most $u-1$ studies have positive effects.
Formally, the composite null hypothesis that at most $u-1$ studies have positive effects is 
$$ H^{u/n}(R): \btheta \notin \mathcal{A}^{u/n}(R) = \cup_{ \boldsymbol{i} \in \Pi(u)} H_{\boldsymbol{i}}^{R},$$

where $\Pi(u)$ is the set of all $n-(u-1)$ tuples from $\{1,\ldots,n\}$, ${\boldsymbol{i}}$ is a tuple in $\Pi(u)$.
Testing the composite null hypothesis is possible by the key observation that it is rejected if and only if for every tuple ${\boldsymbol{i}} := \{i_1,\ldots,i_{n-u+1}\}\in \Pi(u)$, the corresponding  \emph{intersection hypothesis}- $$H^R_{{\boldsymbol{i}}}: \theta_{i_1}\leq 0, \ldots,  \theta_{i_{n-u+1}}\leq 0$$ 
is rejected \citep{Benjamini09}. 
For example, for $u=2$,  at least two  of the $n$ studies have  positive effects if and only if for each of the $n$ subsets of $n-1$ studies,  at least one study has a positive effect. 

A level $\alpha$ test rejects $H^{u/n}(R)$ if for every intersection hypothesis ${\boldsymbol{i}}\in \Pi(u)$, the null $H^R_{{\boldsymbol{i}}}$ is rejected in favor of the alternative that there exists a $j\in {\boldsymbol{i}}$ with $\theta_j>0$ by an $\alpha$ level test.
For each intersection hypothesis  many statistical tests are available that combine the individual $p$-values or test statistics \citep{Loughin04,Futschik19}. 	The preferred test depends on the (unknown) alternative, and there is no single test that dominates all others. Fisher's combining method aggregates the study $p$-values $p^R_{i_1},\ldots,p^R_{i_{n-u+1}}$ with the combining function  $f(p^R_{i_1},\ldots,p^R_{i_{n-u+1}})= -2\sum_{j=1}^{n-u+1} \log p^R_{i_j}$ \citep{fisher1950statistical, littell1971asymptotic}.

Fisher's combining method is popular in various application fields (e.g., genomic research, education, social sciences)  since it has been shown to have excellent power properties \citep{Owen09}. 
It is rarely used in meta-analyses of randomized clinical trials, where the focus is on effect sizes. 
We shall consider the following extension of this combining method, which is useful if the treatment effects are suspected to have mixed signs. In such a case,  a potentially more powerful test is  based on aggregation of the $p$-values that are at most a predefined threshold $t$ \citep{Zaykin02}, 
and the null distribution is adjusted accordingly. For a test at level $\alpha$, \cite{Zaykin02} recommend setting the cut-off threshold at $t=\alpha$ based on empirical investigations. We concur with this recommendation based on our own investigations for our settings, see Figure  \ref{fig-sim2} for details. Our 
test statistic for  the intersection hypothesis $H^R_{i_1,\ldots,i_{n-u+1}}$ is therefore 
\begin{equation}
C_\alpha^R(i_1,\ldots,i_{n-u+1}) = -2\sum_{j\in \{i_1,\ldots,i_{n-u+1} \}} \log \left\lbrace (p^R_{j})^{I[p^R_{j}\leq \alpha]}\right\rbrace.\nonumber	\end{equation}
The null distribution has a simple form. Using the computation method in \cite{Hsu13}, the $p$-value for the intersection hypothesis is 
\begin{equation}\label{eq-truncatedPearson1Sided}
p^R_{\left( i_1,\ldots,i_{n-u+1}\right)}=\sum_{k=1}^{n-u+1}P_{n-u+1,\alpha}\left(k\right) \times\left[1-F_{k}\left\{ -\log\left(\frac{\exp\left(-\frac{	C_\alpha^R(i_1,\ldots,i_{n-u+1})}{2}\right)}{\alpha^{k}}\right)\right\} \right],
\end{equation}
where $F_{k}\left(\cdot\right)$ is to the cumulative gamma distribution with scale parameter equal to one and shape parameter $k$, and $P_{n-u+1,\alpha}\left(\cdot \right)$ is the cumulative Binomial distribution with $n-u+1$ trials and probability of success $\alpha$. 
The $p$-value for $H^{u/n}(R)$ is 
$$r^R (u) = \max_{(i_1,\ldots,i_{n-u+1})\in \Pi(u)} p^R_{\left( i_1,\ldots,i_{n-u+1}\right) }.$$

Since $	C_\alpha^R(i_1,\ldots,i_{n-u+1})$ is monotone in the $p$-values, $r^R(u)$ can be computed efficiently in $O(n\log n)$ computations by sorting the right-sided $p$-values. Then $r^R(u)$ will be the $p$-value of the intersection hypothesis with indices corresponding to the $n-u+1$ largest (i.e., least significant)   $p$-values. Formally, denoting the sorted right-sided $p$-values by $p^R_{(1)}\leq \ldots \leq p^R_{(n)}$,   
$$r^R (u) =   \sum_{k=1}^{n-u+1}P_{n-u+1,\alpha}\left(k\right) \times\left[1-F_{k}\left\{ -\log\left(\frac{\exp\left(-\frac{C_{\alpha}^{R}(u)}{2}\right)}{\alpha^{k}}\right)\right\} \right], $$ where 
$	C_\alpha^R(u) = -2\sum_{j=u}^n \log \left\lbrace \left(p^R_{(j)}\right)^{I\left[p^R_{(j)}\leq \alpha \right]}\right\rbrace.$

The above steps can be straightforwardly adjusted in order to compute the $p$-value for $H^{(u/n)}(L)$, denoted by $r^L(u)$. 

\begin{remark}
	there are many one-sided composite tests to combine
	test statistics from multiple sources, for example, one-sided sum
	test \citep{pocock1987analysis,frick1994maxmin},
	approximate likelihood ratio test \citep{follmann1996simple,tang1989approximate},
	Max test \citep{tarone1981distribution}. 
	For each of these tests, as well as our selected test,  there exists a data generation for which the test is optimal. We  favour Zaykin's combining method, since it handles efficiently $p$-values that are stochastically larger than uniform, a setting which may arise if the study  effects have mixed signs. However, that the $r$-value and confidence lower bounds described below can be applied using any valid one-sided composite test, so researchers can choose their favourite intersection test instead. 
\end{remark}

\subsection{The \textit{r}-value}\label{subsecRV}
The $p$-value of the test with the minimal replicability requirement, i.e. with $u=2$,  is simply referred to as the $r$-value.
The null hypothesis $H^{2/n}$ is true if at most one study has an effect in either direction.
An evidence is \emph{replicable}  if the $r$-value is below the nominal level for the type I error, since then the conclusion is that at least two studies have an effect in the same direction. 

Formally, the	$p$-value for the composite null hypothesis $H^{u/n}: H^{u/n}(R)\cap H^{u/n}(L)$ is $$r(u) = 2\min \{r^R (u),r^L (u) \}.$$ 
\begin{remark}
	With $u=1$, this test reduces to Pearson's test described in \cite{Owen09}, which  is useful for powerful identification of effects that are consistently decreasing or consistently increasing across the $n$ studies. This test has greater power than a test based on Fisher's combining method using two-sided $p$-values when  direction of the treatment effect is consistent across studies, while not requiring us to know the common direction.
\end{remark}

\subsection{Confidence lower bounds for replicability of effect direction}\label{subsec-LB}

In order to establish, with $1-\alpha$ confidence lower bounds on the number of studies with negative effects and the number with positive effects, we test in order $H^{u/n}(L)$ and $H^{u/n}(R)$ for increasing values of $u$ \citep{Heller10}.  
Let $u^L_{\max}$ be the maximal value of $u$ for which $H^{u/n}(L)$ was rejected  at significance level $\alpha/2$, 
$$u^L_{\max} = \arg\max_u\{r^L(u)\leq \alpha/2\}. $$
Then we can conclude with $1-\alpha/2$ confidence, that there are at least $u^L_{\max}$ studies with negative effect. 
Similarly, compute $u^R_{\max} = \arg\max_u\{r^R(u)\leq \alpha/2\}.$ 
Therefore, with $1-\alpha$ confidence, there are at least $u^L_{\max}$ studies with negative effect  and $u^R_{\max}$ studies with a positive effect.

\subsection{Enhancing the meta-analysis report with replicability analysis findings}\label{subsec-enhancing}
By adding the $r$-value we provide an objective measure of the confidence that the finding is not driven by a single study. A result that $r$-value$\leq 0.05$ is useful for strengthening the scientific finding, by  concluding that at least two studies have a treatment effect in the same direction.  However, a result that  $r$-value$> 0.05$  should not be used to undermine the credibility that the treatment effect is replicable, since this result may be  due to lack of power. This result may urge the researcher to seek further evidence for the treatment effect. 

By adding the 95\% lower bounds in each direction, we provide further replicability and consistency evidence.  We say the evidence is   {\it inconsistent} if the lower bounds in each direction are positive: $\min\left\lbrace u^L_{\max} , u^R_{\max} \right\rbrace \geq 1 $. Evidence of \emph{inconsistency} warrants the examination of  why some studies deem the intervention effective and others harmful. We say the evidence {\it supports consistency} in effect direction if (1) $u^L_{\max}
\geq 2$ and $u^R_{\max}=0$, or (2) $u^L_{\max}=0$ and $u^R_{\max}\geq 2$.  
These two additions are useful even when heterogeneity of treatment is modeled by the RE model.

\section{Simulations}\label{sec-simulation}
Simulation studies are carried out to examine the power of, the aforementioned, assumption-free tests for establishing replicability of effect in various settings of interest. 

\subsection{Simulation settings}	

For study $i\in \{1,\ldots,n\}$, the estimated effect size, $\hat \theta_i$, is sampled from the normal distribution with mean $\theta_i$ and standard error $SE_i = \sqrt{1/n_{Ci}+1/n_{Ti}}$, where $n_{Ci}$ and $n_{Ti}$ are the control and treatment group sizes, respectively. 
We examined a wide range of values for $\btheta,n$, and $ (n_{Ci}, n_{Ti})$. 

Since the qualitative conclusions are similar for the various values of $n,n_{Ci}, n_{Ti}$, we display in this section results for $n=8$, with sizes $ \left\lbrace n_{Ci} \right\rbrace = \left\lbrace 22, 210,  26, 192,  60,  38,  53,  15\right\rbrace $ and
$ \left\lbrace n_{Ti} \right\rbrace = \left\lbrace 22, 121,  24, 187,  31,  53,  49,  16\right\rbrace $ 
(these values are similar to those in the example detailed in Figure 8). Simulations for other variations of $n=4,20$ and$\backslash$or $\{ n_{Ci} = n_{Ti} = 25 \}$ are shown in the Supplementary Material.
For the effects vector $\btheta$,
we considered two fundamentally different settings: the fixed effects setting, in which  the effects are fixed to the same value for each data generation, e.g., a common effects value for all the non-null studies; the random effects setting, in which  the effects distribution is the one assumed in the RE model, i.e, for each data generation, i.i.d random samples $\btheta$
are drawn from $\mathcal{N}\left( \mu , \tau^2 \right).$
The number of iterations was set at $10^5$.

\subsection{Simulation results}\label{subsec-sim_results}

\begin{figure}[htpb]
	\centering
	\includegraphics[width=0.9\textwidth]{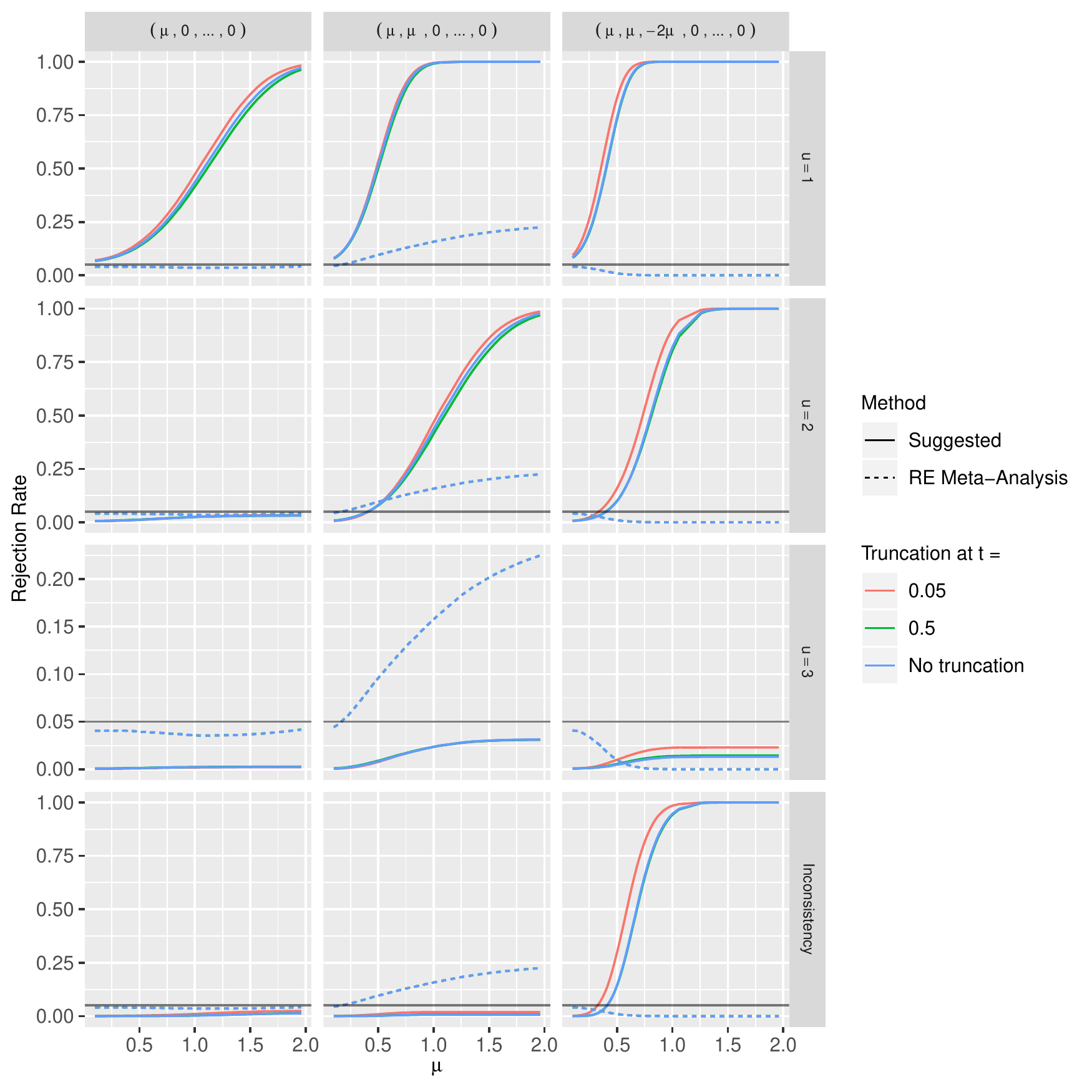} 
	\caption{ Rejection rate as function of the strength of the fixed effects vector characterized by $\mu$. The effects vector $\left( \theta_1,\theta_2, \dots ,\theta_8\right) $ is  indicated at the top of each column. The null hypotheses  examined are: the RE meta-analysis null  (dashed), the global null $H^{1/n}$ (row 1), the minimal replicability null $H^{2/n}$ (row 2), $H^{3/n}$ (row 3), no lack of consistency (row 4). Except for the RE meta-analysis null, the test statistics  use products of  truncated $p$-values at most: 0.05 (solid red); 0.5 (solid green); or 1 (solid blue). The horizontal solid line is the 0.05 significance level of the test. The effect estimates $\hat{\theta}_j$ are sampled from the normal distribution with mean $\theta_j$ and standard deviation $SE_j$.   The replicability null hypothesis, $H^{2/n}$, is true in the first column  and false otherwise. The inconsistent setting is in the left column. 
	}\label{fig-sim2}
\end{figure}

We find that the power of the replicability analysis can be much higher than of the meta-analysis in many realistic FE and RE settings.
Figure 	\ref{fig-sim2} shows three fixed effects settings for the eight studies: a treatment effect only in a single study, the same effect in only two studies, and three studies with effect but with mixed signs. We examined three truncation values for the test statistic, and found that for all values of $u$, as well as for detecting inconsistency, the best truncation value is at the nominal level for the type I error. The advantage  over $t=0.5$ or $t=1$ (i.e., no truncation) is especially large in the setting with mixed signs. Therefore, from henceforth we only consider truncation at $t=0.05$ (our nominal type I error level). As expected, the power to reject $H^{2/n}$ when it is false is lower than that of rejecting $H^{1/n}$, but it increases to one as the signal strengthens in two studies in the same direction. The power to detect inconsistency increases to one as well when the signs are mixed.  Arguably, the meta-analysis that will be carried out for these fixed effects data generations is a RE meta-analysis because of the non-negligible heterogeneity. 
The RE meta-analysis rejects the null hypothesis of no overall treatment effect at most 5\% of the time when the treatment effect is present in only one study, thus providing better protection against the danger of concluding there is a treatment effect based on a single study than the FE meta-analysis. 
The power to detect an overall signal with the RE model is very low also in the other two settings:  at most  $15\%$ when a treatment effect in the same direction is present in two studies; at most  $5\%$ when the treatment effect is inconsistent. However, in these two settings of the middle and right columns the replicability  can be established with power increasing to one as the absolute value of the treatment effects increase.

\begin{figure}[htpb]
	\centering
	\includegraphics[page = 1, width=0.6\textwidth, height= 0.3\textheight]{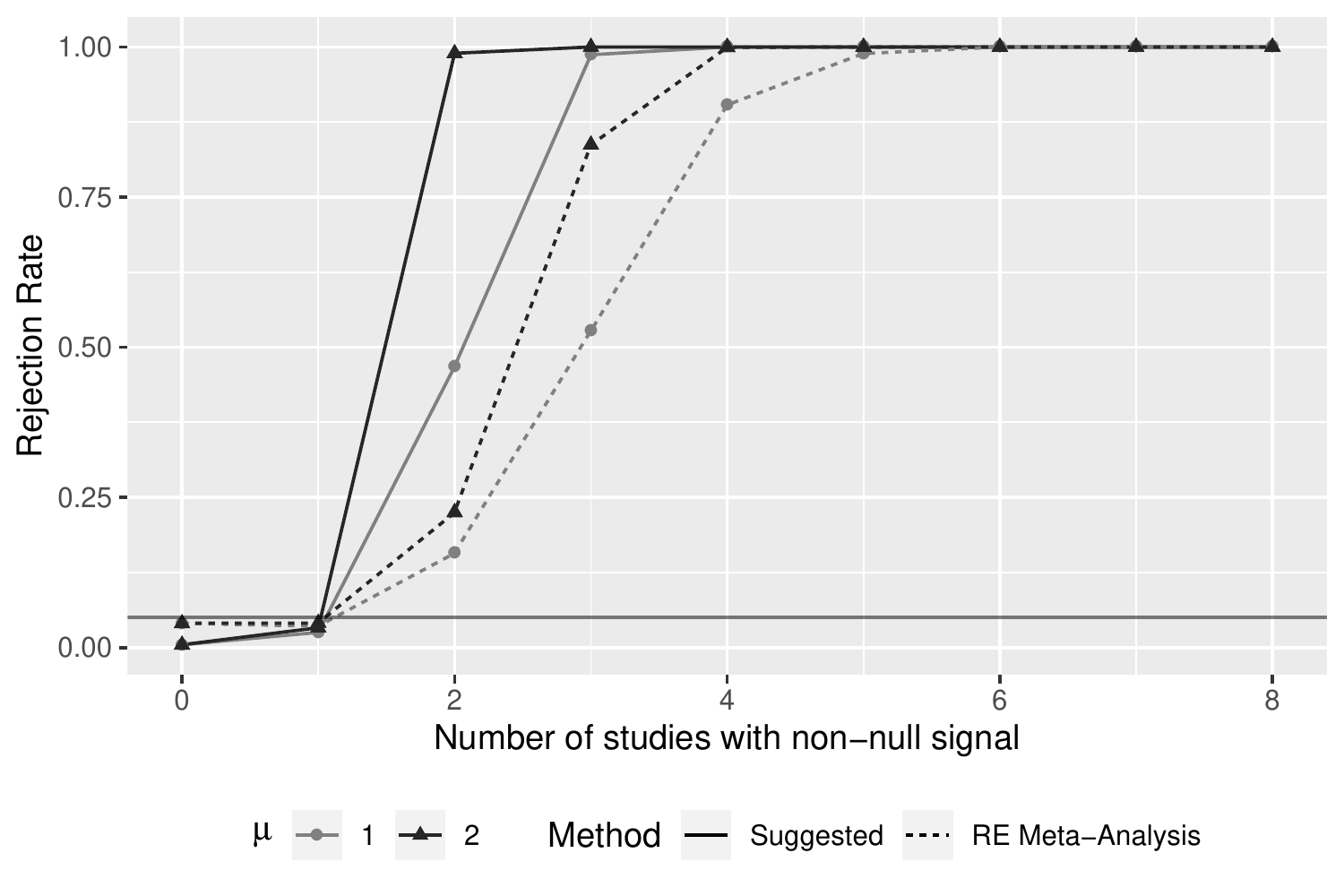} 
	\caption{ Rejection rate versus the number of nonnull studies with a common treatment effect, with the RE meta-analysis test (dashed) or with the no replicability test of $H^{2/n}$ (solid).    The curves with: circles, triangles, and squares, have treatment effect  value of one, two, and three, respectively.  The replicability null hypothesis, $H^{2/n}$,  is true when the number of studies is zero or one, and false otherwise. The horizontal solid line is the 0.05 significance  level of  the test. The product of truncated $p$-values are truncated at $t=0.05$. 
	}\label{fig-simFE}
\end{figure}

Figure \ref{fig-simFE} shows fixed effects settings where the non-null treatment effect is common, and the number of studies with this common effect increases from zero to eight.
The rejection rate of the repliability null hypothesis $H^{2/n}$ (i.e., with minimal replicability requirement) is far greater than  that of the RE meta-analysis null hypothesis when the number of non-null studies is at least two, but the gap diminishes as the number of non-null studies increases.

\begin{figure}[htpb]
	\centering
	\includegraphics[width=0.8\textwidth, height= 0.3\textheight]{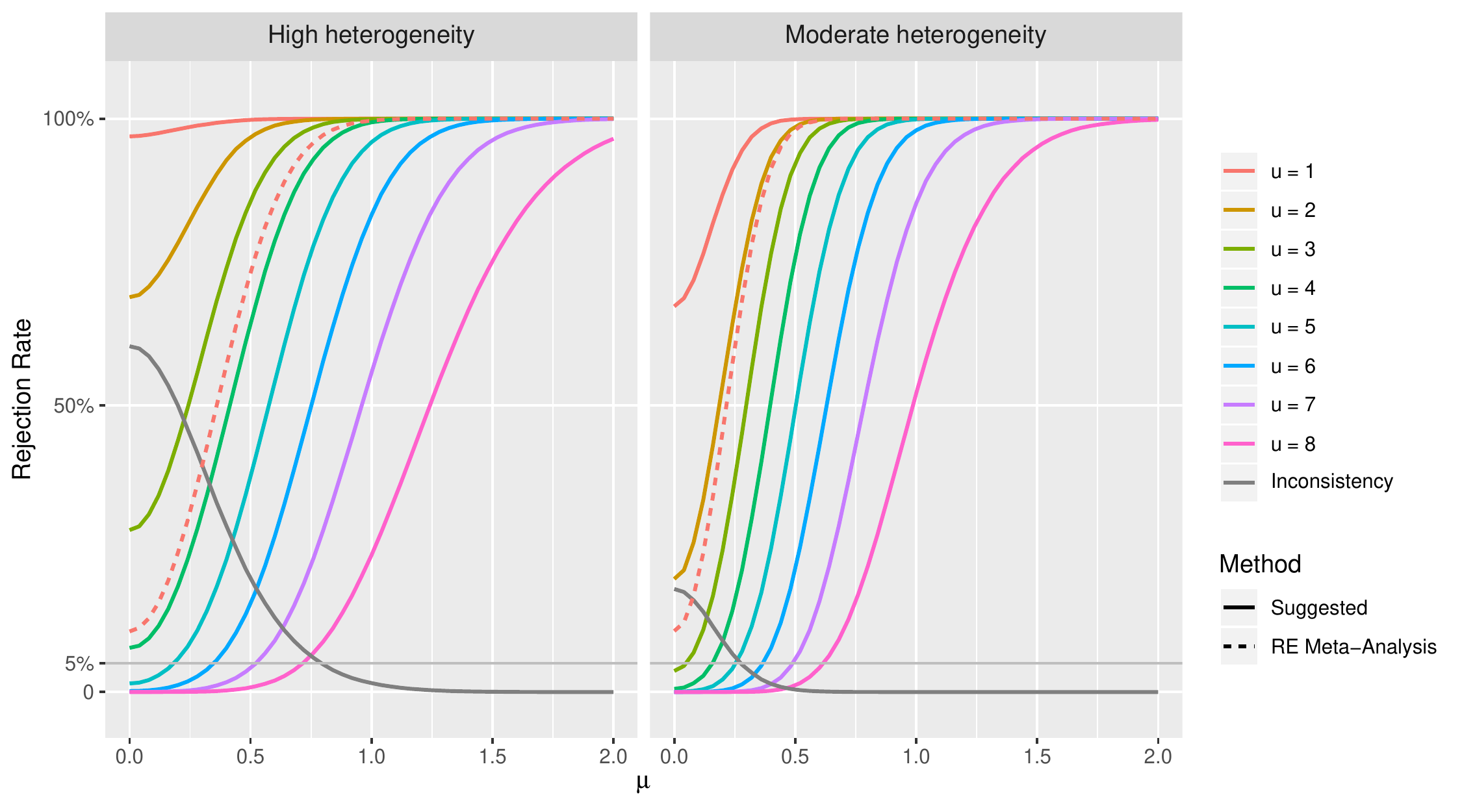} 
	\caption{ Rejection rate as a function of the overall treatment effect $\mu$ in data generated according to the RE model. The hypotheses tests examined are: the RE Meta-analysis null that $\mu=0$ (dashed), and  $H^{u/n}$ for $u=1,\ldots,8$ (solid). For study $j=1,\ldots,8$, treatment effect estimate $\hat \theta_j$ is sampled form the normal distribution with mean $
		\theta_j$ and standard deviation $SE_j$. The treatment effect $
		\theta_j$ is itself sampled independently from the normal distribution with mean $\mu$ and standard deviation $\tau$. The value of $\tau$ is chosen so the estimated heterogeneity is around 70\% in the left panel, and around 50\% in the right panel. The product of truncated $p$-values are truncated at $t=0.05$. 
	}\label{fig-sim3}
\end{figure}

Figure \ref{fig-sim3} shows random effects settings with high and  moderate heterogeneity. For this data generation, the effect sizes are non-zero with probability one. Therefore, the minimal replicability null hypothesis is never true. 
The power for discovering minimum replicability, i.e. the test of $H^{2/8}$, is greater in all settings than the power to discover the overall effect by the RE meta-analysis. As expected, the power decreases as $u$, the minimum number of studies with effect in the same direction we want to discover, increases.  

When the data is generated according to the RE model, the probability of inconsistency, i.e., of having at least one positive and one negative treatment effect, increases as the overall mean approaches zero and as the heterogeneity increases. For $\theta_i\sim N(\mu, \tau^2)$, the probability to have an inconsistent configuration is $1-\Phi(\mu/\tau)^n - \left\lbrace 1-\Phi(\mu/\tau)\right \rbrace^n$, which is $1-\left(\frac12\right)^{n-1}\approx 1$ when $\mu=0$. We declare (and thus detect) inconsistency if the lower bound for both the decreasing effect and the increasing effect is at least one. 
The probability of detecting that the effects are inconsistent   in the setting considered in Figure \ref{fig-sim3} reached $\approx 60\%$ in the setting with $\mu=0$ and high heterogeneity, and deteriorated quickly as $\mu$ increased. Potentially more powerful tests for detecting inconsistency are available \citep{gail1985testing, piantadosi1993comparison}, but these tests do not provide lower bounds on the number of studies with effect in each direction.

	\section{Case studies from the Cochrane library}\label{sec-examples}

We provide examples of meta-analyses in the breast cancer domain for which we can, and cannot, claim replicability.  For each example, we  report the $r$-value (as described in \S~\ref{subsecRV} with $\alpha=0.05$) and the  95\% confidence lower bounds on the number of studies with effect in each direction (as described in \S~\ref{subsec-LB}). 
Moreover, we provide recommendations on  how to incorporate these new analyses in the   abstract and forest plots of Cochrane reviews.

\begin{figure}[ht]
	\centering
	\includegraphics[width=\textwidth]{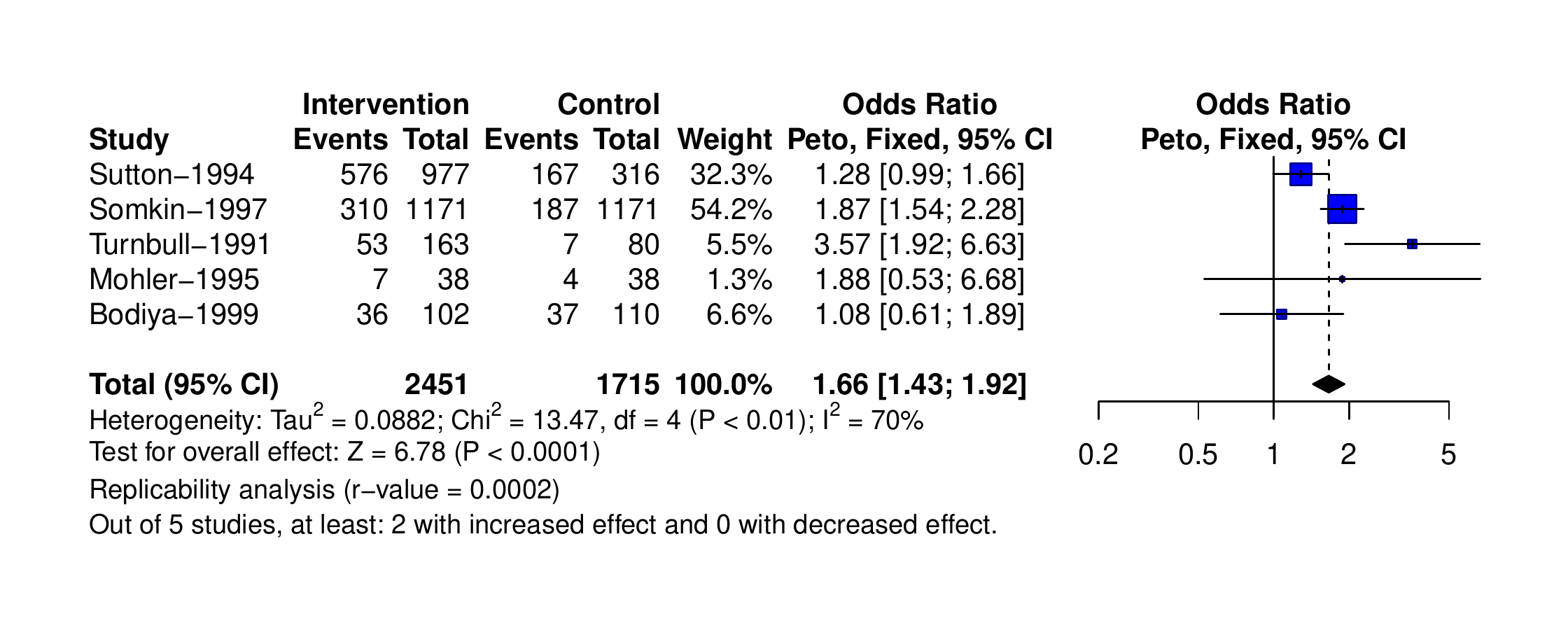}
	\caption{In review CD002943, the effect of mammogram invitation on attendance during the following 12 months. The evidence towards replicability is strong: the $2-out-of-5$   $r-value = 5\times10^{-4}$; the 95\% lower bound on the number of studies with increased effect, relative to 1, is 2 studies.  }\label{fig-CD002943}
\end{figure}

The first example is based on a FE meta analysis in review CD002943 (Figure~\ref{fig-CD002943}.
The primary objective of this review  was to assess the effectiveness of different strategies for increasing the participation rate of women invited to community breast cancer screening activities or mammography programs.   In this meta-analysis, the effect of sending  invitation letters was examined in five studies.
The authors main result is that: "The odds ratio in relation to the outcome, 'attendance in response to the mammogram invitation during the 12 months after the invitation, was 1.66 ($95\%$ CI 1.43 to 1.92)".
As was previously suggested in  §\ref{intro} and §\ref{subsec-LB}, we suggest adding the $r$-value and lower confidence bounds on the number of studies, as follows: "The evidence towards an increased response rate was replicable, with  $r-value=0.0002$. Moreover, with 95\% confidence, we can conclude that  at least two  studies had an increased effect."

\begin{figure}[ht]
	\centering
	\includegraphics[width=\textwidth]{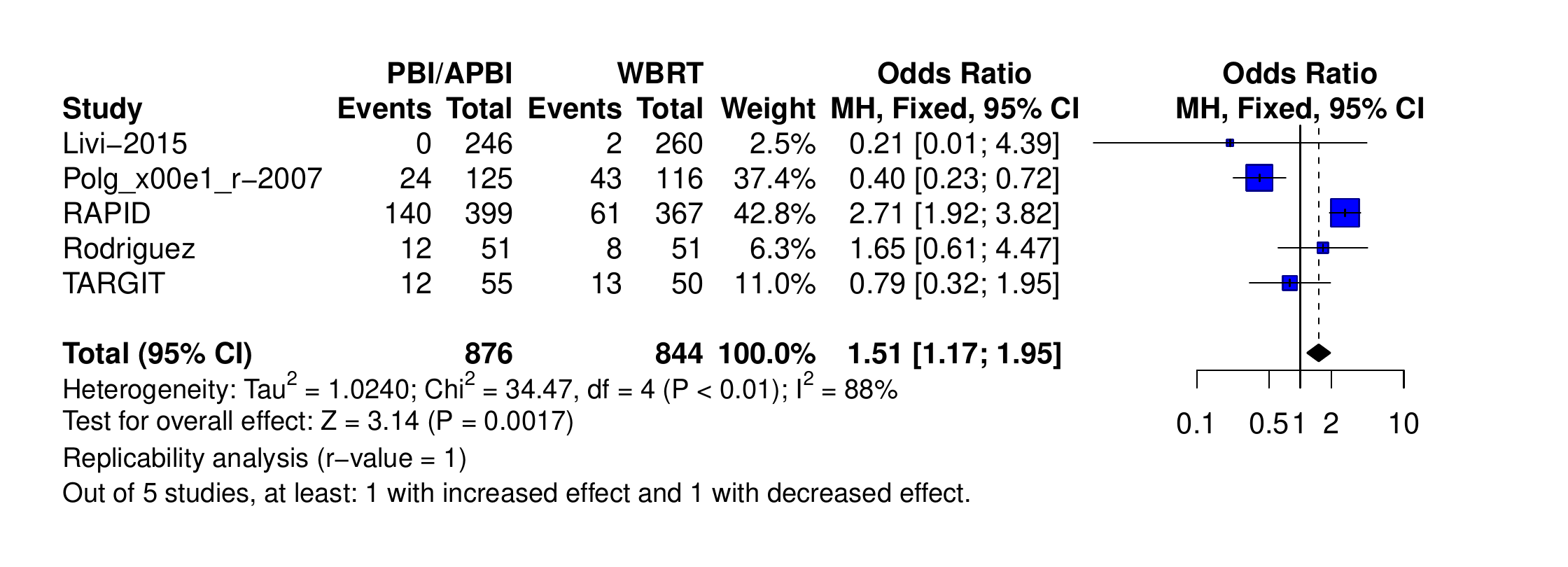} 
	\caption{ In review CD007077, the effect of PBI/APBI versus  WBRT on Cosmesis. There is no evidence towards replicability, $r(2)$-value $\approx$1. At least one study report in increased effect relative to 1, and one more reports a decreasing effect, thus we warn against inconsistency across the studies combined.} \label{fig-CD007077}
\end{figure}

The second example is based on a FE meta analysis in review CD007077 (Figure~\ref{fig-CD007077} 
) regarding after breast-conservation therapy for early-stage breast cancer. 
The primary objective of this review was to assess the effectiveness of partial breast irradiation (PBI) or accelerated partial breast irradiation (APBI), compared to the conventional or hypo-fractionated whole breast radiotherapy (WBRT). 
The  primary outcome was Cosmesis. The meta-analysis overall  effect is significant, with a 95\% CI entirely to the right of the null value, despite the fact that  the two largest studies report  conflicting significant effects.
Therefore, the authors write as a main result that "Cosmesis (physician-reported) appeared worse with PBI/APBI (odds ratio (OR) 1.51, $95\%$ CI 1.17 to 1.95, five studies, 1720 participants, low-quality evidence)". 
We suggest adding 
"We cannot rule out the possibility that this result is critically based on a single study ( $r-value= 1$). Moreover, the results are inconsistent, since with 95\% confidence, we conlclude that at least one study had an increased effect and at least one study had a decreased effect."
In both examples, the narrow 95\% CI does not convey information about the lack or existence of replicability of the effect across studies, but our additional inference does.

\begin{figure}[ht]
	\centering
	\includegraphics[width=\textwidth]{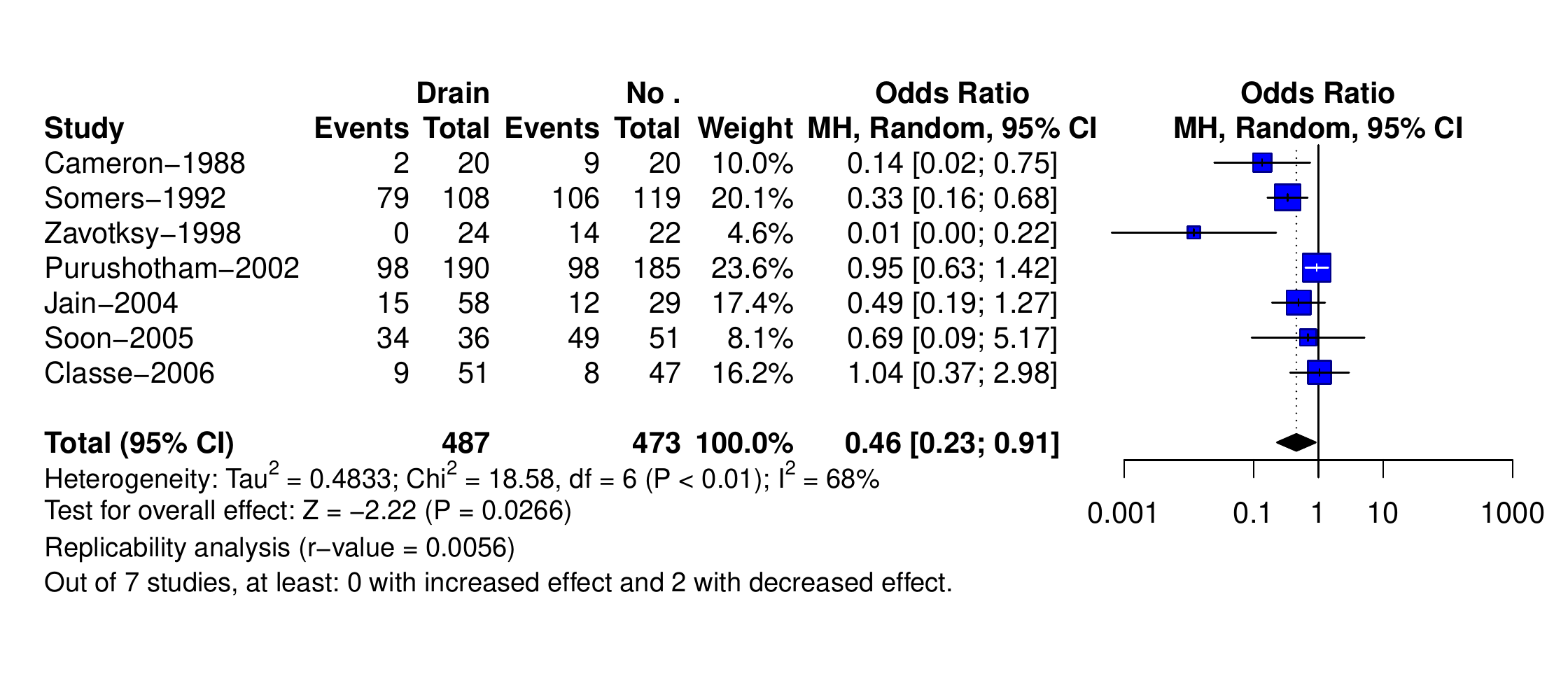} 
	\caption{ In review CD006823, the effects of wound drainage on Seroma formation. The evidence is consistent: the $2-out-of-7$ $r-value = 0.0012$; there is a  decreased effect (relative to 1) in at least 2 out of 7 studies, and no study with increased effect, with 95\% confidence. } \label{fig-CD006823}
\end{figure}

The third  example is based on a RE meta analysis in review CD006823 (Figure~\ref{fig-CD006823}), where the meta-analysis finding was statistically significant. The authors examine the effects of wound drainage after axillary dissection for breast carcinoma on the incidence of post-operative Seroma formation. 
The authors write  "The OR for Seroma formation was 0.46 ($95\%$ CI 0.23 to 0.91, P = 0.03) in favor of a reduced incidence of Seroma in participants with drains inserted." To this, we suggest adding our additional analysis as follows:  "The evidence towards a decreased effect was replicable ($r$-value = 0.0056). Moreover, with 95\% confidence, we conclude that at least two studies had a reduction effect, with no indication of incosistency."

\begin{figure}[ht]
	\centering
	\includegraphics[width=\textwidth]{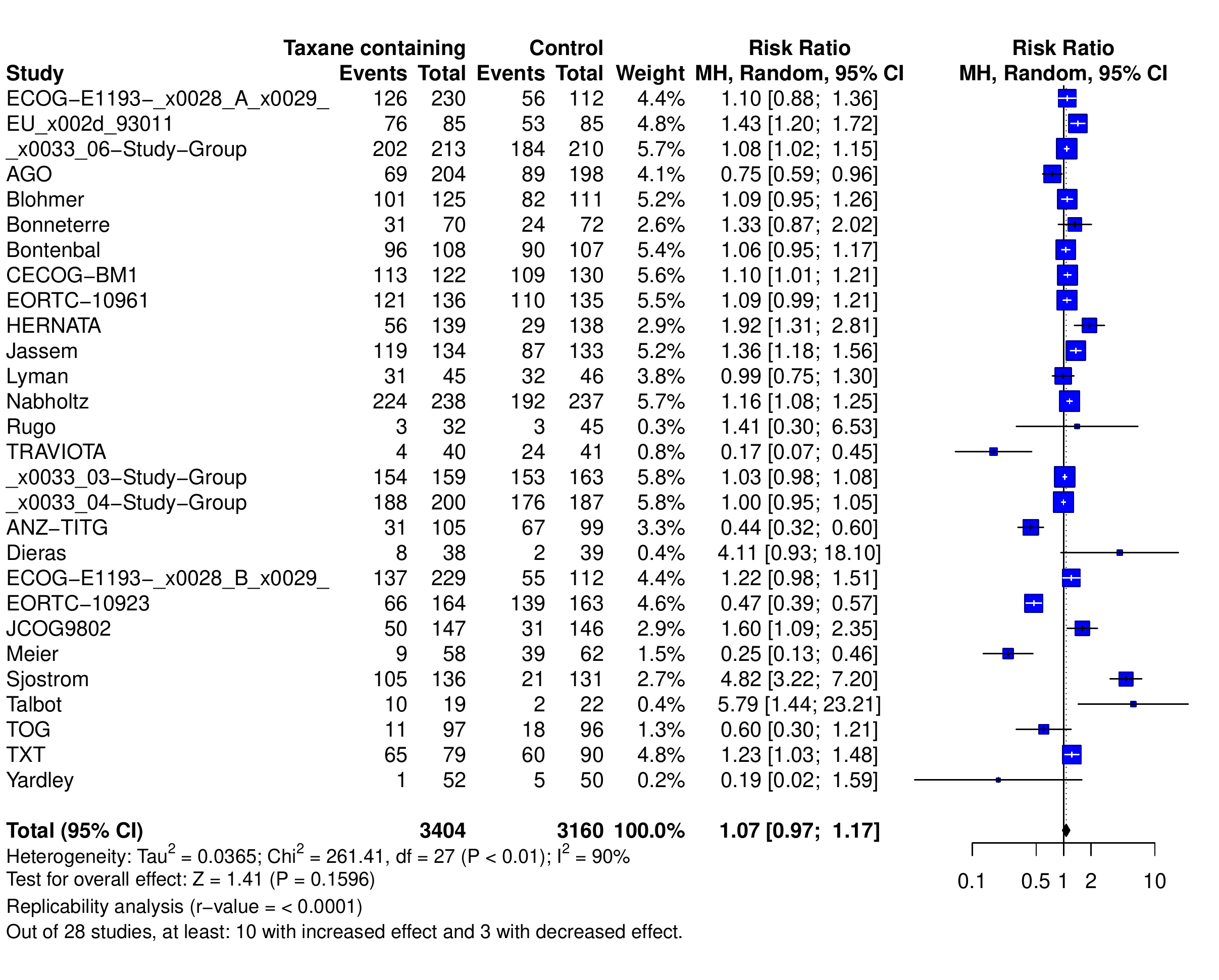}
	\caption{In review CD003366, the effect of  chemotherapy regimens on  Leukopaenia. The evidence towards both an increased and a decreased effect is strong: the 2 out of 28 $r$-value is $<0.0001$; the 95\% lower bound on the number of studies with increased and decreased risk relative to 1 is 10 and 3, respectively.} 
	\label{fig:CD003366}
\end{figure}

The fourth example is based on a RE meta-analysis in review CD003366
(Figure~\ref{fig:CD003366}). The authors compare chemotherapy regimens on overall effect in
Leukopaenia. Pooling 28 studies, the RE meta-analysis fails to
declare any significant difference between regimens, due to the
highly-significant yet contradicting results. The authors write: "Overall,
there was no difference in the risk of Leukopaenia (RR 1.07; 95\% CI 0.97 to
1.17; P = 0.16; participants = 6564; Analysis 5.2) with significant
heterogeneity across the studies (I2 = 90\%; P $<$ 0.00001)".  We
suggest adding: "There is inconsistent evidence for the direction of
effect: an increased risk of Leukopaenia in at least ten studies and a decreased
risk in at least three studies (with 95\% confidence)."

\section{Replicability assessment in the breast cancer domain }\label{sec-extentOfproblem}

We took all the updated Cochrane Collaboration systematic reviews in  breast cancer domain. Our eligibility criteria were as follows: (a) the review included forest plots; (b) at least one fixed-effect primary outcome was reported as significant at the .05 level, which is the default significant level used in Cochrane Reviews; (c) the meta-analysis of at least one of the primary outcomes was based on at least three studies  (d) there was no reporting in the review of unreliable/biased primary outcomes or poor quality of available evidence, and (e) the data is available for download. We consider as primary outcomes the outcomes that were defined as primary by the review authors. If none were defined we selected the most important findings from the review summaries and treated the outcomes for these findings as primary.
In the breast cancer domain 62 updated (up to February 2018) reviews were published by the Cochrane Breast Cancer Group in the Cochrane library, out of which we analyzed 23  reviews that met our eligibility criteria (16, 12 , 5 , 2 and 4 reviews was excluded due reasons a, b, c, d and e respectively). Out of the 23 eligible reviews, in 8 reviews we had enough evidence to establish replicability (i.e., an $r$-value at most 0.05) for all the primary outcomes with meta-analysis $p$-values at most 0.05.

	\begin{figure}[ht]
	\centering
	\includegraphics[width=\textwidth]{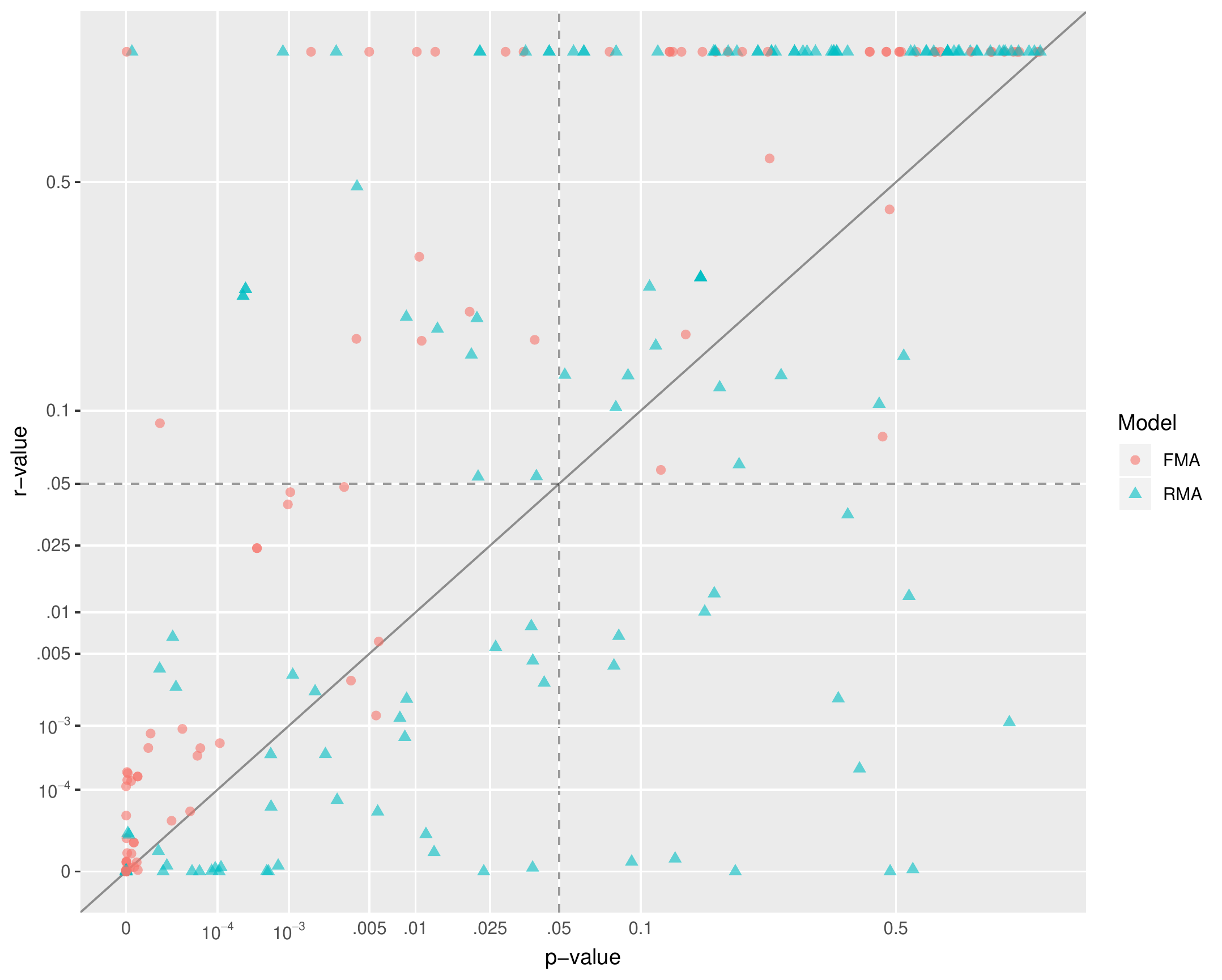}
	\caption{ $p$-values versus $r$-values, after the quartic root ( power of $1/4$) transformation, for each of the 245 primary outcomes analyzed with the fixed-effect model (red circles), or the random effects model (blue triangles). The axes show the matching values on the original scale. Color darkness increases according to the number of overlapping results.The solid line is the diagonal line of $45^{o}$. }
	\label{fig:PvaluesVsRvalues}
\end{figure}

We analyzed a total of 245  primary outcomes contributed by the eligible systematic reviews of which 105  were FE meta-analyses, as reported by the authors. Out of the 70 outcomes with a  statistically significant FE $p$-value, 57 were replicable ($r$-value $\leq 0.05$). For the 57 replicable findings, we rule out the danger that the discovery is entirely driven by one study. Thus, the evidence on the treatment effect is more trustworthy. The importance of detecting replicability for trusting the evidence in favour (or harm) of a treatment is manifest, for example, in the FDA requirement for at least two studies finding an effect \citep{mdi1998guidance}.  

For the 245 primary outcomes, Figure \ref{fig:PvaluesVsRvalues} shows the $r$-values versus the meta-analysis $p$-values,  and Table \ref{tab} summarizes the consistency evidence. 
As expected, among the non-significant outcomes the fraction of studies supporting  consistency is smaller than among the significant outcomes. Ten inconsistent outcomes were detected, nine of which were analyzed via RE model by the authors, 
warranting  further research into why the effects are inconsistent across studies.

	\begin{table}
	\small\sf\centering
	\caption{The evidence towards consistency and inconsistency in the 245 meta-analyses, for   significant (at the 0.05 level) meta-analysis outcomes (column 2) and non-significant meta-analysis outcomes (column 1).}\label{tab}
	\begin{tabular}{|c | c| c|}
		\hline
		& non-significant meta-analysis $p-$value & significant meta-analysis $p-$value  \\ 
		\hline
		\hline
		Evidence supports consistency &   7 &  96 \\ 
		Evidence inconsistent &  10 &   8 \\ 
		Not enough evidence &  96 &  28 \\ 
		\hline
	\end{tabular}
\end{table}

\section{ The special case of a nonnull common effect }\label{sec-rvalueFE}

Assuming that the nonnull studies have a common effect, a powerful test statistic for the intersection hypothesis is the one used by the FE model. Specifically, $\theta_i\in \{0, \theta\} $ for $i=1,\ldots,n$. The FE model alternative is that all studies have a common effice, i.e.,  $\theta_i = \theta$ for all $i$.

For a subset of studies $\{i_1,\ldots, i_{n-u+1} \}$, the  estimated common effect and standard error are  $$\widehat{\theta}_{(i_1,\ldots,i_{n-u+1})} = \frac{\sum_{k=1}^{n-u+1} \frac{\hat \theta_{i_k} }{\widehat{SE}^2_{i_k}}}{\sum_{k=1}^{n-u+1} \frac{1}{\widehat{SE}^2_{i_k}}}, \quad
\widehat{SE}_{(i_1,\ldots,i_{n-u+1})} = \frac{1}{\sqrt{\sum_{k=1}^{n-u+1} \frac{1}{\widehat{SE}^2_{i_k}}}}.$$ 
The $p$-value for the intersection hypothesis $H^R_{i_1,\ldots,i_{n-u+1}}$ is $$p^R_{(i_1,\ldots,i_{n-u+1})} = 1-\Phi\left(\frac{\widehat{\theta}_{(i_1,\ldots,i_{n-u+1})}}{\widehat{SE}_{(i_1,\ldots,i_{n-u+1})}} \right).   $$ 
The $p$-value for  $H^L_{i_1,\ldots,i_{n-u+1}}$  is $p^L_{(i_1,\ldots,i_{n-u+1})} =1-p^R_{(i_1,\ldots,i_{n-u+1})}$. 
The $p$-value for $H^{u/n}(X)$, $X\in \{L,R\}$ is 
$$r_{FE} ^X (u) = \max_{(i_1,\ldots,i_{n-u+1})\in \Pi(u)} p^X_{\left( i_1,\ldots,i_{n-u+1}\right) }.$$
The 	$p$-value for  $H^{u/n}: H^{u/n}(R)\cap H^{u/n}(L)$ is 
$$r_{FE}(u) = 2\min \{r_{FE}^R (u),r_{FE}^L (u) \}.$$
Intuitively, the $r$-values should be larger than the meta-analysis $p$-value since a stronger scientific claim is made by rejecting $H^{u/n}$ than by rejecting the FE meta-analysis null hypothesis. We formalize this in the following proposition (see %\ref{supp:Appendix_proof}
 in supplementary for the proof).

\begin{proposition}\label{thm-FE}
	Let $p = 2\min\{p^L_{(1,\ldots,n)}, p^R_{(1,\ldots,n)} \} $ be the FE meta-analysis $p$-value. Then, if $\theta_i\in \{0, \theta\} $ for $i=1,\ldots,n$, for $u\in \{2, \ldots,n\} $: $p<r_{FE}(u)$; if $\widehat{\theta}_{(1,\ldots,n)}<0$,  $p^L_{1,\ldots,n}<r_{FE} ^L (u)$;  if $\widehat{\theta}_{(1,\ldots,n)}>0$,  $p^R_{1,\ldots,n}<r_{FE} ^R (u)$.
\end{proposition}

See Supplementary for proof; for further simulations see figures %\ref{supp:fig-sim2_extensions} 
and %\ref{supp:fig-sim3_extensions}
 in the supplementary.

Fig.~\ref{fig-simFE_FRA} shows the sensitivity of the FE model to the setting with exactly one nonnull study.

\begin{figure}[htpb]
	\centering
	\includegraphics[width=0.55\textwidth, height= 0.4\textheight]{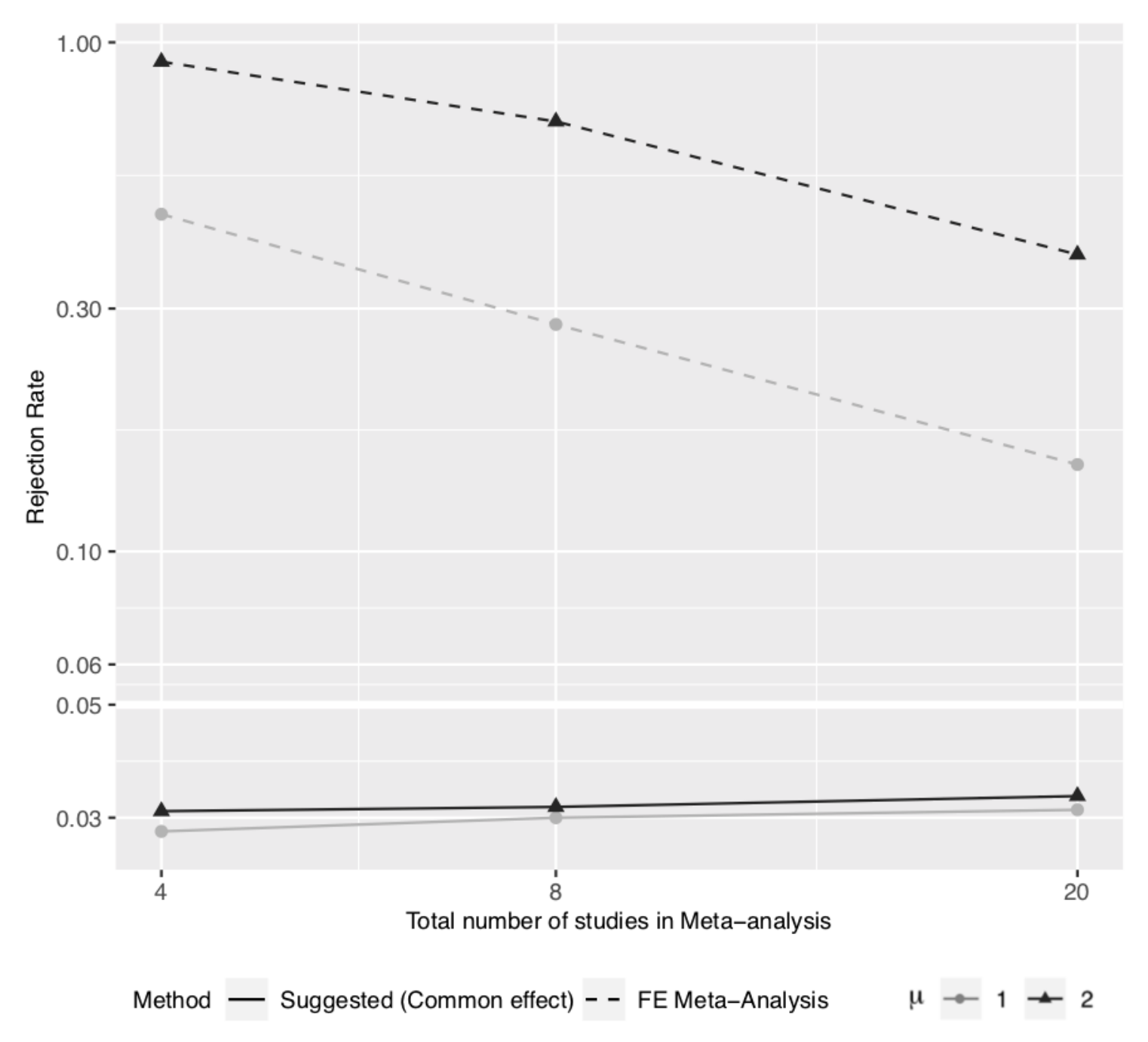} 
	\caption{ Rejection rate versus the total number of studies, with the FE meta-analysis test (dashed), or the test of the replicability null hypothesis, $H^{2/n}$, with the common-effect assumption (solid). Only a single study has an effect, with effect size one (circles) or two (triangles). Therefore, the FE meta-analysis null hypothesis is false but  $H^{2/n}$,  is true.  The rejection rate is much higher than the 0.05 nominal level for the test of the FE meta-analysis null hypothesis, and it is below 0.05  for $H^{2/n}$.
	}\label{fig-simFE_FRA}
\end{figure}

\section{Discussion}\label{sec-discuss} 

We provided examples, mainly from the Cochrane library, to demonstrate the  benefit from complementing the meta-analysis with a report  of the $r$-value and lower bounds on the number of studies with increasing and decreasing effect; although, it should be clear that the methods we offer can be used in any meta-analysis.
We recommend adding the quantified information to the forest plot of the meta-analysis as a convention. 
We specifically encourage adding  it to the two-page abstract of the Cochrane systematic review, which is a standalone document (published in MEDLINE) that briefly reports the main results and author's conclusions.

Seemingly, it may be thought that if the number of studies is large, the meta-analysis cannot be driven by one outlying study. However, we found four fairly large fixed-effect analyses, with 17, 11, 9 and 7  studies, for which the meta-analysis $p$-value was significant unlike the $r$-value. 

Establishing replicability for 2 studies out of 4  is a stronger statement than out of 20. This is where introducing $u_{max}$ bounds offer flexible view on replicability. 
The appropriate size of $u_{max}$ relative to $n$ is a question to be explored in each scientific discipline.
Nevertheless, the $r$-value representing a minimal requirement, 
is valuable when $n$ 
is small or large: a significant $r$-value when pooling a small number of studies reflects  strong evidence towards replicability of effects; a non-significant $r$-value for numerous studies salvages from unfounded results.

High heterogeneity can be a result of numerous reasons, whether inevitable or not. Regardless of its source, heterogeneity may lead to studies having opposite signs for estimated effect sizes which makes pooling the estimated effects meaningless. An unperceiving mean effect does not disclose the contradictions between the different studies, but rather diminish the valuable underlying findings. 
When pooling a big number of studies 
would be tolerable to observe both negative and positive effect estimates, regardless of the true magnitudes of effect or between study variance. Assuming the true effect is positive, it would be expected that a small portion of the studies report negative estimates. A statistically founded calculation of the bounds $u^{R}_{max}, u^{L}_{max}$ reflects the strength of each registered signal.

The suggested complementary replicability analysis gives insight into the effects consistency. For example, we see from Table \ref{tab} that among the outcomes with a non-significant random effects $p$-value,  we have evidence supporting consistency in $7$  and inconsistency in $10$  of the 113 meta-analyses with a non-significant RE meta-analysis $p$-value.

 High heterogeneity can also appear when the dispersion of the estimated effects contributes most of the overall variance, as argued by \cite{borenstein2017basics}, leading to non-significant RE meta-analysis $p$-value. 
If, in spite of that, we observe that the effects have consistent signals, it can still lead to a non-significant $p$-value. 

If the evidence supports consistency, then the overall meta-analysis CI informs the user about the effect size in an appropriate manner. If inconsistency is established, then the overall meta-analysis  CI 
may not provide clinically meaningful information
for assessing whether the intervention is beneficial or harmful, unless there is an explanation why some studies should be excluded from the treatment evaluation. It is, however,  possible to incorporate the effect size into the replicability analysis by considering tests of certain effects instead of no effect. For example, in order to identify whether the effect change is at least $\Delta$ in at least $u$ studies,  testing can proceed as described in \S~\ref{subsec-NRhyp} with zero  replaced by $\Delta$ in the hypotheses definitions. Similarly, an upper bound on the "positive" effect size can be conveniently found being the maximal value of $\Delta > 0$ such that the corresponding $r-$value is at most $\alpha/2$; a lower bound on the minimal "negative" effect can be determined as well. Bear in mind that the upper bound of $\Delta$ is only meaningful if $r^R-$value$<\alpha/2$; the same applies to the lower bound on $\Delta$ and $r^L-$value.

The only caveat is that meta-analysis is prone
to publication bias, where only significant results (at $p-value \leq .05$ )
are published. The Cochrane reviews are known to be careful
during their search for eligible studies, avoiding as much as possible this problem.
In other areas, where this may not be feasible, using conditional p-values rather than the raw ones in the procedures may circumvent the problem (with unfortunate loss of some power.) More concretely, we can use the result of \citep{zhao2019multiple} that showed that for one-sided tests in a one-dimensional exponential family, the conditional distribution of the $p$-value divided by $\alpha$, given ghat the $p$-value was at most $\alpha$, is stochastically larger than the uniform (0,1) distribution for any $\alpha$. Thus limiting our inference only to the set of hypotheses with $p$-values at most 0.05 (for example), and using a valid combining function on these $p$-values divided by 0.05, provides a valid inference immune at least to some forms of publication bias (specifically, that studies are published only if they are below a certain threshold that is at least 0.05).

\section{Supporting information}
An R package implementing the methods proposed in this paper is now available for download at CRAN, under the name 'metarep' (\url{https://cran.r-project.org/web/packages/metarep/index.html}). 
R-codes for generating the reported simulations  and reproducing the examples are  
available on GitHub (\url{https://github.com/IJaljuli/r-value}). 
% Supplemental file includes  (1) results of simulations like in \S\ref{sec-simulation} for $n=4,20$ and both equal and unequal group sizes, and (2) proof for Proposition \ref{thm-FE}. 

\section*{Acknowledgments}
The authors thank  Ian Pan for his help with extracting and processing data from Cochrane forest plots; 
David Steinberg for useful comments on an earlier version of this manuscript; and Daniel Yekutieli for useful discussions about the methodology.

\section*{Funding}
This research was supported by the Israeli Science Foundation [grant 1049/16 (RH)]; Joint US-NSF and US-Israel BSF [grant no. 2016746 (YB, IJ)]; The Baroness Ariane de Rothschild Women's Doctoral grant[(IJ)];
and U.S. Agency for Healthcare Research and Quality [grant no. 1R03HS025840 (OAP)].


\begin{thebibliography}{99}
	
	\bibitem[Anzures-Cabrera and Higgins, 2010]{AnzuresCarbera10}
	Anzures-Cabrera, J. and Higgins, J. P.~T. ({2010}).
	\newblock {Graphical displays for meta-analysis: An overview with suggestions
		for practice}.
	\newblock {\em {Research Synthesis Methods}}, {1}({1}):{66--80}.
	
	\bibitem[Bax et~al., 2006]{Bax06}
	Bax, L., Yu, L.-M., Ikeda, N., Tsuruta, H., and Moons, K.~G. ({2006}).
	\newblock {Development and validation of mix: comprehensive free software for
	meta-analysis of causal research data}.
	\newblock {\em {BMC Medical Research Methodology}}, {6}({1}):{50}.
	
	\bibitem[Benjamini and Heller, 2008]{conj}
	Benjamini, Y. and Heller, R. ({2008}).
	\newblock {Screening for partial conjunction hypotheses}.
	\newblock {\em {Biometrics}}, {64}({4}):{1215--1222}.
	
	\bibitem[Benjamini et~al., 2009]{Benjamini09}
	Benjamini, Y., Heller, R., and Yekutieli, D. ({2009}).
	\newblock {Selective inference in complex research}.
	\newblock {\em {Philosophical Transactions of the Royal Society A-Mathematical Physical and Engineering Sciences}}, {367}({1906}):{4255--4271}.
	
	\bibitem[Borenstein et~al., 2009]{Borenstein09}
	Borenstein, M., Hedges, L., and Higgins, J. ({2009}).
	\newblock {Rothstein hr.}
	\newblock {\em {Introduction to Meta-Analysis}}.
	
	\bibitem[Borenstein et~al., 2017]{borenstein2017basics}
	Borenstein, M., Higgins, J. P.~T., Hedges, L.~V., and Rothstein, H.~R.
	({2017}).
	\newblock {Basics of meta-analysis: I-2 is not an absolute measure of
		heterogeneity}.
	\newblock {\em {Research Synthesis Methods}}, {8}({1}):{5--18}.
	
	\bibitem[Collins and Tabak, 2014]{collins2014policy}
	Collins, F.~S. and Tabak, L.~A. ({2014}).
	\newblock {NIH plans to enhance reproducibility}.
	\newblock {\em {Nature}}, {505}({7485}):{612--613}.
	
	\bibitem[Deeks et~al., 2019]{Deeks19}
	Deeks, J.~J., Higgins, J.~P., Altman, D.~G., and Group, C. S.~M. ({2019}).
	\newblock {Analysing data and undertaking meta-analyses}.
	\newblock {\em {Cochrane Handbook for Systematic Reviews of Interventions}},
	{pages 241--284}.
	
	\bibitem[Fisher, 1936]{fisher1935design}
	Fisher, R.~A. ({1936}).
	\newblock {Design of experiments}.
	\newblock {\em{ BMJ}}, {1}({3923}):{554--554}.
	
	\bibitem[Fisher, 1950]{fisher1950statistical}
	Fisher, R.~A. ({1950}).
	\newblock {Statistical methods for research workers}.
	\newblock {\em {Statistical Methods for Research Workers}}, ({llth ed. revised}).
	
	\bibitem[Follmann, 1996]{follmann1996simple}
	Follmann, D. ({1996}).
	\newblock {A simple multivariate test for one-sided alternatives}.
	\newblock {\em {Journal of the American Statistical Association}},
	{91}({434}):{854--861}.
	
	\bibitem[Frick, 1994]{frick1994maxmin}
	Frick, H. ({1994}).
	\newblock {A maxmin linear test of normal means and its application to Lachin's
		data}.
	\newblock {\em {Communications in Statistics-Theory and Methods}},
	{23}({4}):{1021--1029}.
	
	\bibitem[Futschik et~al., 2019]{Futschik19}
	Futschik, A., Taus, T., and Zehetmayer, S. ({2019}).
	\newblock {An omnibus test for the global null hypothesis}.
	\newblock {\em {Statistical Methods in Medical Research}},
	{28}({8}):{2292--2304}.
	
	\bibitem[Gail. and Simon, 1985]{gail1985testing}
	Gail., M. and Simon, R. ({1985}).
	\newblock {Testing for qualitative interactions between treatment effects and
		patient subsets}.
	\newblock {\em {Biometrics}}, {41}({2}):{361--372}.
	
	\bibitem[Gaudino et~al., 2020]{JAMA_ioi200073}
	Gaudino, M., Hameed, I., Farkouh, M.~E., Rahouma, M., Naik, A., Robinson,
	N.~B., Ruan, Y., Demetres, M., Biondi-Zoccai, G., Angiolillo, D.~J.,
	Bagiella, E., Charlson, M.~E., Benedetto, U., Ruel, M., Taggart, D.~P.,
	Girardi, L.~N., Bhatt, D.~L., and Fremes, S.~E. ({2020}).
	\newblock {Overall and cause-specific mortality in randomized clinical trials
	comparing percutaneous interventions with coronary bypass surgery: A
	meta-analysis}.
	\newblock {\em{ JAMA Internal Medicine}}.
	
	\bibitem[Heller, 2011]{Heller10}
	Heller, R. ({2011}).
	\newblock {Discussion of "multiple testing for exploratory research" by
	\textit{JJ Goeman} and \textit{A. Solari}}.
	\newblock {\em {Statistical Science}}, {pages 598--600}.
	
	\bibitem[Higgins, 2011]{Higgins11}
	Higgins, J. ({2011}).
	\newblock {Cochrane handbook for systematic reviews of interventions. version
	5.1. 0 [updated march {2011}]. the cochrane collaboration}.
	\newblock {\em {www.cochrane-handbook.org}}.
	
	\bibitem[Higgins and Thompson, 2002]{higgins2002quantifying}
	Higgins, J. and Thompson, S. ({2002}).
	\newblock {Quantifying heterogeneity in a meta-analysis}.
	\newblock {\em {Statistics in Medicine}}, {21}({11}):{1539--1558}.
	\newblock {International symposium on methodological issues in systematic
		reviews and meta-analysis, University of Oxford, Oxford, England, Jul. 03-05,
		2000}.
	
	\bibitem[Higgins et~al., 2009]{higgins2009re}
	Higgins, J. P.~T., Thompson, S.~G., and Spiegelhalter, D.~J. ({2009}).
	\newblock {A re-evaluation of random-effects meta-analysis}.
	\newblock {\em {Journal of the Royal Statistical Society Series A-Statistics in
			Society}}, {172}({1}):{137--159}.
	
	\bibitem[Hsu et~al., 2013]{Hsu13}
	Hsu, J.~Y., Small, D.~S., and Rosenbaum, P.~R. ({2013}).
	\newblock {Effect modification and design sensitivity in observational
		studies}.
	\newblock {\em {Journal of the American Statistical Association}},
	{108}({501}):{135--148}.
	
	\bibitem[Kafkafi et~al., 2005]{Kafkafi05}
	Kafkafi, N., Benjamini, Y., Sakov, A., Elmer, G., and Golani, I. ({2005}).
	\newblock {Genotype-environment interactions in mouse behavior: A way out of
		the problem}.
	\newblock {\em {Proceedings of the National Academy of Sciences of the United
			States of America}}, {102}({12}):{4619--4624}.
	
	\bibitem[Littell and Folks, 1971 ]{littell1971asymptotic}
	Littell, R.~C. and Folks, J.~L. ({1971}).
	\newblock {Asymptotic optimality of Fisher's method of combining independent
		tests}.
	\newblock {\em {Journal of the American Statistical Association}}.
	
	\bibitem[Loughin, 2004]{Loughin04}
	Loughin, T. ({2004}).
	\newblock {A systematic comparison of methods for combining p-values from
		independent tests}.
	\newblock {\em {Computational Statistics \& Data Analysis}},
	{47}({3}):{467--485}.
	
	\bibitem[McNutt, 2014]{mcnutt2014journals}
	McNutt, M. ({2014}).
	\newblock {Journals unite for reproducibility}.
	
	\bibitem[MDI and Drug, 1998]{mdi1998guidance}
	MDI, M. D.~I. and Drug, D. P. I.~D. ({1998}).
	\newblock {Guidance for industry}.
	\newblock {\em {Center for Drug Evaluation and Research (CDER)}}, {1000}.
	
	\bibitem[{National Academies of Sciences, Engineering, and Medicine},
	2019]{national2019reproducibility}
	{National Academies of Sciences, Engineering, and Medicine} ({2019}).
	\newblock {\em {Reproducibility and replicability in science}}.
	\newblock {National Academies Press}.
	
	\bibitem[Nuzzo, 2014]{nuzzo2014scientific}
	Nuzzo, R. ({2014}).
	\newblock {Scientific method: statistical errors}.
	\newblock {\em {Nature News}}, {506}({7487}):{150--152}.
	
	\bibitem[Owen, 2009]{Owen09}
	Owen, A.~B. ({2009}).
	\newblock {Karl Pearson's meta-analysis revised}.
	\newblock {\em {Annals of Statistics}}, {37}({6B}):{3867--3892}.
	
	\bibitem[Panagiotou and Trikalinos, 2015]{panagiotou2015effect}
	Panagiotou, O.~A. and Trikalinos, T.~A. ({2015}).
	\newblock {On effect measures, heterogeneity, and the laws of nature}.
	\newblock {\em {Epidemiology}}, {26}({5}):{710--713}.
	
	\bibitem[Piantadosi and Gail, 1993]{piantadosi1993comparison}
	Piantadosi, S. and Gail, M. ({1993}).
	\newblock{ A comparison of the power of two tests for qualitative interactions.}
	\newblock {\em {Statistics in Medicine}}, {12}({13}):{1239--1248}.
	
	\bibitem[Pocock and Tsiatis, 1987]{pocock1987analysis}
	Pocock, S.~J., G. N.~L. and Tsiatis, A.~A. ({1987}).
	\newblock {The analysis of multiple endpoints in clinical trials}.
	\newblock {\em {Biometrics}}, {43}({3}):{487--498}.
	
	\bibitem[Riley et~al., 2011]{riley2011interpretation}
	Riley, R.~D., Higgins, J. P.~T., and Deeks, J.~J. ({2011}).
	\newblock {Interpretation of random effects meta-analyses}.
	\newblock {\em {British Medical Journal}}, {342}.
	
	\bibitem[science collaboration, 2015]{open2015estimating}
	Science Collaboration, O. ({2015}).
	\newblock {Estimating the reproducibility of psychological science}.
	\newblock {\em {Science}}, {349}({6251}):{4716}.
	
	\bibitem[Tang and Geller, 1989]{tang1989approximate}
	Tang, D.~I., G.~C. and Geller, N.~L. ({1989}).
	\newblock {An approximate likelihood ratio test for a normal mean vector with
		nonnegative components with application to clinical trials}.
	\newblock {\em {Biometrika}}, {76}({3}):{577--583}.
	
	\bibitem[Tarone, 1981]{tarone1981distribution}
	Tarone, R.~E. ({1981}).
	\newblock {On the distribution of the maximum of the logrank statistic and the
		modified Wilcoxon statistic}.
	\newblock {\em {Biometrics}}, {37}({1}):{79--85}.
	
	\bibitem[Wang and Owen, 2019]{wang2019admissibility}
	Wang, J. and Owen, A.~B. ({2019}).
	\newblock {Admissibility in partial conjunction testing}.
	\newblock {\em {Journal of the American Statistical Association}},
	{114}({525}):{158--168}.
	
	\bibitem[Zaykin et~al., 2002]{Zaykin02}
	Zaykin, D., Zhivotovsky, L., Westfall, P., and Weir, B. ({2002}).
	\newblock {Truncated product method for combining P-values}.
	\newblock {\em {Genetic Epidemiology}}, {22}({2}):{170--185}.
	
	\bibitem[Zhao et~al., 2019]{zhao2019multiple}
	Zhao, Q., Small, D.~S., and Su, W. ({2019}).
	\newblock {Multiple testing when many p-values are uniformly conservative, with	application to testing qualitative interaction in educational interventions}.
	\newblock {\em {Journal of the American Statistical Association}}, {114}({527}):{1291--1304}.
	
\end{thebibliography}
\end{document}